\documentclass{article}
\usepackage[centertags]{amsmath}

\begin{document}

\title{Nonsymmetric entropy I: basic concepts and results }

\author{ Cheng-Shi Liu \\Department of Mathematics\\Daqing Petroleum Institute\\Daqing 163318, China
\\Email: chengshiliu-68@126.com}

 \maketitle

\begin{abstract}
 A new concept named nonsymmetric entropy which generalizes the concepts of
  Boltzman's entropy and shannon's entropy, was introduced. Maximal nonsymmetric entropy principle
  was proven. Some important distribution laws were derived naturally from maximal
   nonsymmetric entropy principle.\\

\end{abstract}
\section{introduction}
Entropy which measures the uncertain degree of information is an
important basic concept in statistic physics and information theory.
In Ref.[1] and its references, entropy is discussed from many
aspects. In present paper, I give a new entropy named nonsymmetric
entropy which measures the average value of the auxiliary and
probability two kinds of information to an event. I prove the
corresponding maximal nonsymmetric entropy principle. Some
interesting distribution laws can be derived naturally from this
principle.

\section{Basic conceptions and results}

Firstly we give some concepts in the following.

 \textbf{Definition 1}:The auxiliary
information of an event $x_i$ is defined by
\begin{equation}
A(x_i)=-\ln(\beta_i),
\end{equation}
where $\beta_i$ is auxiliary information parameter.

\textbf{Definition 2}: Total information of an event $x_i$ is
\begin{equation}
It(x_i)=I(x_i)+A(x_i)=-\ln p(i)-\ln \beta_i=-\ln(\beta_i p(i)).
\end{equation}

\textbf{Definition 3}: we define a function
\begin{equation}\label{WE}
S_m(p(1),\cdots,p(m))=-\sum _{i=1}^m p(i)\ln(\beta_ip(i)),
\end{equation}
where $\beta_i>0, i=1,\cdots, m$, are nonsymmetric parameters. We
call the function $S_m$ the nonsymmetric entropy.

\textbf{Remark 1}: if we take $\beta_i\equiv1$, we get the Shannon's
entropy, thus nonsymmetric entropy generalizes the concept of
Shannon's entropy.

It is obvious that nonsymmetric entropy measures the expect
information of the total information of all events. Because we
consider auxiliary information so that we can describe event in a
more right way. At the same time,  we can use nonsymmetric
parameters $\beta_i$ to derive some important distribution laws
which include Zipf's law. In particular, if we take
\begin{equation}
\beta_i=i^\alpha,
\end{equation}
then the corresponding nonsymmetric entropy becomes
\begin{equation}
S=-\sum _{i=1}^m p(i)\ln(i^\alpha p(i)),
\end{equation}
we can call it Zipf's entropy. We have the following result:

\textbf{Theorem 1}: If $p(i)$ satisfies the following Zipf's
distribution law
\begin{equation}
p(i)=\frac{p(1)}{i^\alpha},
\end{equation}
then the Zipf's entropy takes the maximum.

 \textbf{Corollary 1}. For Zipf's law, we have $S_{m+1}>S_m$, that is, the nonsymmetric entropy
 is increase as $m$ increasing.

From the above theorem it is easy to see that the Zipf's law can be
derived naturally from nonsymmetric entropy under some special
auxiliary parameters. We don't prove this theorem at the present
time, in fact we have the following more general result.

\textbf{Theorem 2}: If  $\{p(1),\cdots,p(m)\}$ satisfies the
following distribution
\begin{equation}
p(i)=\frac{1}{\beta_i\sum_{i=1}^m \frac{1}{\beta_i}},
\end{equation}
then the nonsymmetric entropy $S_m$ takes the maximum
\begin{equation}
S_m=-\ln p(1)=\ln \sum_{i=1}^m \frac{1}{\beta_i}.
\end{equation}\\

\textbf{Proof}: Instituting $p(m)=1-p(1)-\cdots-p(m-1)$ into the
Eq.(\ref{WE}) and setting its differential to zero yields
\begin{equation}\label{Eu}
\frac{\partial S_m}{\partial
p(i)}=-\ln\frac{\beta_ip(i)}{\beta_m(1-p(1)-\cdots-p(m-1))}=0,
 \ \ i=1,\cdots,m-1,
\end{equation}
that is,
\begin{equation}\label{AE}
p(1)+\cdots+(1+\frac{\beta_i}{\beta_m})p(i)+\cdots+p(m-1)=1,
 \ \ i=1,\cdots,m-1.
\end{equation}
Solving the equations system (\ref{AE}), we obtain
\begin{equation}
p(i)=\frac{1}{\beta_i\Sigma_{j=1}^m\frac{1}{\beta_j}}=\frac{\beta_1}{\beta_i}p(1),
\ \i=1,\cdots,m,
\end{equation}
where $p(1)=\frac{1}{\beta_1\Sigma_{j=1}^m\frac{1}{\beta_j}}$.
Denote $A_k=(a_{ij})_{k\times k},
a_{ij}=\frac{\delta_{ij}}{p(i)}+\frac{1}{p(m)}$, since
$\det{A_k}=\frac{1}{p(m)}\Sigma_{i=1}^k\frac{1}{p(1)\cdots
\widehat{p(i)}\cdots p(k)}$, where the hat means the corresponding
item disappear, so from $a_{ij}=-\frac{\partial^2S_m}{\partial
p(i)\partial p(j)}=\frac{\delta_{ij}}{p(i)}+\frac{1}{p(m)}$, we know
that the matrix $A=(a_{ij})_{(m-1)\times(m-1)}$ is a positive
defined matrix. Thus the distribution
$p(i)=\frac{1}{\beta_i\sum_{i=1}^m \frac{1}{\beta_i}}$ maximize
the nonsymmetric entropy. The proof is completed.\\

\textbf{Remark 2}: If we take $ \beta_i=i^\alpha$, then we have
\begin{equation}
p(i)=\frac{1}{\Sigma_{j=1}^m\frac{1}{j^\alpha}}=\frac{p(1)}{i^\alpha},
\ \i=1,\cdots,m,
\end{equation}
in particular, we take $\alpha\simeq1$, this is just the Zipf's law
in linguist. If take $\beta_i=(i+\gamma)^\alpha$, then we give
Mandelbrot's law. If we take other values of $\beta_i$, we will give
other distribution law. Thus the key is to choose suitable auxiliary
information parameters $\beta_i$, this is a problem need to study
deeply.

\textbf{Remark 3}: Using maximal nonsymmetric entropy principle in
Section 3, we can give a simple proof for theorem 2.\\

We consider the continuous case in the following.

 \textbf{Definition
4}: For continuous case, nonsymmetric entropy is defined
\begin{equation}
S(\rho)=-\int \rho(x)\ln\{\beta(x)\rho(x)\}\mathrm{d}x,
\end{equation}
where $\beta(x)$ is auxiliary information parameter function,
$\rho(x)$ is probability density of event.

In order to solve maximal nonsymmetric entropy distribution, we can
use variant method. Under some constrains conditions, we use
lagrange multiply method to do this thing. We give several example
in the following to illustrate our method.

\textbf{Theorem 2}: Assume $\int x\rho(x)\mathrm{d}x=\mu$, we then
its maximal nonsymmetric entropy distribution is
\begin{equation}
\rho_0(x)=\frac{1}{\beta(x)}\exp(1-\lambda_1-\lambda_2 x),
\end{equation}
where $\lambda_1$ and $\lambda_2$ satisfy two constrain conditions
$\int \frac{1}{\beta(x)}\exp(1-\lambda_1-\lambda_2 x)\mathrm{d}x=1$
and $\int \frac{x}{\beta(x)}\exp(1-\lambda_1-\lambda_2
x)\mathrm{d}x=\mu$.

\textbf{Proof}: Make a auxiliary functional
\begin{equation}
F(\rho, \lambda_1, \lambda_2)=-\int
\rho(x)\ln\{\beta(x)\rho(x)\}\mathrm{d}x+\lambda_1(\int
\rho(x)\mathrm{d}x-1)+\lambda_2(\int x\rho(x)\mathrm{d}x)-\mu).
\end{equation}
We have
\begin{equation}
\delta F=\int \{\lambda_1+\lambda_2x-\ln(\beta(x)\rho(x))-1\}\delta
\rho(x)\mathrm{d}x,
\end{equation}
form $\delta F=0$, we solve out as follows:
\begin{equation}
\rho_0(x)=\frac{1}{\beta(x)}\exp(1-\lambda_1-\lambda_2 x),
\end{equation}
where $\lambda_1$ and $\lambda_2$ satisfy two constrain conditions
$\int \rho_0(x)\mathrm{d}x=1$ and $\int x\rho_0(x)\mathrm{d}x=\mu$.

\textbf{Theorem 3}: Assume $\int x\rho(x)\mathrm{d}x=\mu$ and $\int
x^2\rho(x)\mathrm{d}x=\sigma^2$, we then its maximal nonsymmetric
entropy distribution is
\begin{equation}
\rho_0(x)=\frac{1}{\beta(x)}\exp(1-\lambda_1-\lambda_2
x-\lambda_3x^2),
\end{equation}
where $\lambda_1, \lambda_2$ and $\lambda_3$ satisfy three constrain
conditions $\int \rho_0(x)\mathrm{d}x=1$ , $\int
x\rho_0(x)\mathrm{d}x=\mu$ and $\int
x^2\rho_0(x)\mathrm{d}x=\sigma^2$.

\textbf{Proof}: it is similar with the proof of theorem 2.

\section{Maximal nonsymmetric entropy principle}

We generalize the maximal entropy principle in information theory to
the case of nonsymmetric entropy.

\textbf{Definition 5}: Denote $\Lambda$ be a class of probability
density functions, if $\rho_0\in\Lambda$, such that
\begin{equation}
S(\rho_0)=max\{S(\rho):\rho\in\Lambda\},
\end{equation}
then $\rho_0$ is called maximal nonsymmetric entropy distribution,
and $S(\rho_0)$ maximal nonsymmetric entropy.

\textbf{Theorem 4} (\textbf{maximal nonsymmetric entropy
principle}): $\Lambda$ is a fixed class of probability density
functions, if there exists $\rho_0\in\Lambda$ such that
\begin{equation}
-\int \rho(x)\ln\{\beta(x)\rho_0(x)\}\mathrm{d}x=S_0
\end{equation}
is a constant which is irrelative to $\rho(x)$ for every $\rho$,
then $\rho_0(x)$ is maximal nonsymmetric entropy distribution, and
$S(\rho_0)=S_0$ is maximal nonsymmetric entropy. For discrete case,
this theorem is also right.

\textbf{Proof}: for arbitrary $\rho\in\Lambda$, we have
\begin{eqnarray}
S(\rho)=-\int \rho(x)\ln\{\beta(x)\rho(x)\}\mathrm{d}x=-\int
\rho(x)\ln\{\beta(x)\rho_0(x)\frac{\rho(x)}{\rho_0(x)}\}\mathrm{d}x\cr
=-\int \rho(x)\ln\{\beta(x)\rho_0(x)\}\mathrm{d}x-\int
\rho(x)\ln\frac{\rho(x)}{\rho_0(x)}\mathrm{d}x\cr \leq-\int
\rho(x)\ln\{\beta(x)\rho_0(x)\}\mathrm{d}x=S_0,
\end{eqnarray}
then $S_0$ is maximal nonsymmetric entropy. Since $S(\rho_0)=S_0$,
so $\rho_0(x)$ is maximal nonsymmetric entropy distribution. The
proof is completed.

\textbf{Corollary 2}: If $\int \frac{1}{\beta(x)}\mathrm{d}x<
\infty$, we have
\begin{equation}
\rho_0(x)=\frac{C}{\beta(x)}
\end{equation}
is maximal nonsymmetric entropy distribution, where $C=\frac{1}{\int
\frac{1}{\beta(x)}\mathrm{d}x}$.

If we take $\beta(x)=x^\alpha, \alpha>1$, and assume the arrange of
random variable $X$ is $(k,+\infty)$, then maximal nonsymmetric
entropy distribution is
\begin{equation}
\rho_0(x)=\frac{k^{1-\alpha}}{\alpha-1}x^{1-\alpha},
\end{equation}
it is just the power law distribution in continuous case. If there
are some constrains we will get other  distributions similar with
them in theorem 2 and theorem 3. On the other hand, we can easily
use the maximal nonsymmetric entropy principle to give new proof s
to theorems 2 and 3.

\section{Discussions}

The above results suggest that the nonsymmetric entropy is a rather
fundamental concept that will play an important role in some fields.
Perhaps the meaning of the nonsymmetric entropy needs a reasonable
explanation. It is different with Shannon's entropy $S=-\sum_{i=1}^m
p(i) \ln p(i)$ in some aspects. For example, if $p(j)=1$ and others
zeroes, then $S=0$, but $S_m=-\ln\beta_j$, this implies that there
exist some kind of uncertainty in some superficial reliable events
under the nonsymmetric entropy. Other deep meanings of nonsymmetric
entropy need more studies. Since the important roles of Boltzman's
entropy and Shannon's entropy in thermodynamics and information
theory respectively, we hope that maximal nonsymmetric entropy
principle can play a suitable role in corresponding fields.

For example, Zipf's law (\cite{Zi})which states that the frequency
of a word decays as a power law of its rank, is regarded as a basic
hypothesis with no need for explanation in recent models of the
evolution of syntactic communication(\cite{NW}). As an empirical
law, Zipf's law is the most fundamental fact in quantitative
linguistics, its meaning is still an open problem which has been
tried to explain from several aspects of its origins(\cite{Ma,Si,
NB,PT,CO}. It is necessary to find a suitable mechanism for Zipf's
law. In this paper, Zipf's law is derived naturally by maximizing
the nonsymmetric entropy when auxiliary parameter take some special
values. It is at least need to consider seriously the meaning of
those results. I will continue to study the theory and applications
of nonsymmetric entropy.

\end{document}